\documentclass[11pt,a4paper]{article}
\makeatletter \oddsidemargin -.1in \evensidemargin -.1in
\newcommand{\singlespacing}{\let\CS=\@currsize\renewcommand{\baselinestretch}{1}\tiny\CS}
\newcommand{\doublespacing}{\let\CS=\@currsize\renewcommand{\baselinestretch}{1.35}\tiny\CS}
\usepackage[utf8]{inputenc}
\usepackage{blindtext}
\usepackage{hyperref}
\usepackage{graphics}
\usepackage{amsmath}
\usepackage{amssymb}
\usepackage{float, epsfig, floatflt, here}
\usepackage{cite}
\usepackage{longtable}
\usepackage{graphicx}
\usepackage{epsfig}
\usepackage{epstopdf}
\usepackage{pdfpages}
\usepackage[T1]{fontenc}
\usepackage{qcircuit}
\usepackage{adjustbox}
\usepackage{subcaption}
\usepackage{braket}

\title {\textbf{\Large Blind Remote Implementation of Arbitrary Two-Qubit Unitary Operation via Pre-Shared Entanglement}}
\author{ \textbf{Plaban Saha}$^{1}$ \thanks{e-mail : plabansaha.iiest@gmail.com }, \textbf{ Binayak S. Choudhury}$^{1,2}$ \thanks {e-mail : binayak@math.iiests.ac.in}
\\ $^{1}$ Department of Mathematics,
 Indian Institute of Engineering Science and Technology,\\ Shibpur
 B. Garden, Howrah - 711103,
West Bengal, India\\
$^{2}$ Department of Mathematics and Applied Mathematics, University of Johannesburg,\\ Auckland Park, Johannesburg- 2006, South Africa\\}
\begin{document}
\date{}
\maketitle

\begin{abstract}
We propose a protocol for the remote implementation of an arbitrary two-qubit unitary operation using pre-shared entanglement. In our protocol, Bob possesses an unknown two-qubit state, while Alice implements the required operations using calibrated black-box devices. Alice does not need to know the parameters of these operations, and Bob's unknown state never leaves his laboratory. The protocol is based on the Euler decomposition of single-qubit unitary operations and the Cartan decomposition of two-qubit unitary operations. First, we construct protocols for the remote implementation of the single-qubit operation $U = R_z(\gamma)R_x(\beta)R_z(\alpha)$ using pre-shared Bell states. We then construct a protocol for the two-qubit operation $R_{zz}(\theta)$ using a single pre-shared Bell state and extend it to $R_{xx}(\theta)$ and $R_{yy}(\theta)$ through local unitary transformations. These elementary protocols are finally combined to implement an arbitrary two-qubit unitary operation. A third party, Charlie, determines the required decomposition and provides Alice with the corresponding black-box devices before the protocol begins. Thus, neither Alice nor Bob requires the classical description of the unitary operation.
\end{abstract}


\section{Introduction}

Quantum teleportation is one of the fundamental protocols of quantum information processing. In the standard quantum teleportation protocol introduced by Bennett et al., an unknown quantum state is transferred from a sender to a spatially separated receiver using a pre-shared entangled state and classical communication \cite{bennett}. Thus, the primary objective of quantum teleportation is the transfer of quantum-state information from one location to another.

A different problem arises when the objective is to implement a quantum operation on a state held by a distant party. This problem is known as remote implementation of quantum operations (RIO), or quantum operation teleportation (QOT). Huelga et al. \cite{huelga1} introduced the concept of quantum remote control for an arbitrary unitary operation $U$. In their general setting, Bob possesses an unknown state $\ket{\psi}$, while Alice has access to the operation $U$. The implementation can be realized by first teleporting $\ket{\psi}$ from Bob to Alice using pre-shared entanglement, applying $U$ to the teleported state at Alice's side, and then teleporting the resulting state $U\ket{\psi}$ back to Bob. Consequently, the operation is physically applied to a copy of the target state located at Alice's side, rather than to Bob's original physical qubit. For an arbitrary operation available as a classical black box, this general setting requires the resources associated with bidirectional state teleportation.

The resource requirement can be reduced when the operation belongs to a restricted class. Huelga, Plenio, and Vaccaro \cite{huelga2} showed that certain continuous sets of single-qubit operations, including arbitrary rotations about a fixed axis, admit remote implementation protocols using fewer resources than bidirectional state teleportation. In such protocols, the target state can remain at the receiver's location while the effect of the operation is implemented remotely. Thus, the operation can be realized on the receiver's target qubit without transferring the target state to the operation holder. Subsequent studies have further investigated remote implementation of quantum operations in different settings \cite{wang,qot0,qot1,qot2,qot3,qot4,qot5,qot6,qot7,qot8,qot9,qot10,qot11}.

More recently, Shi et al. \cite{Shi2025} proposed a protocol for the remote implementation of an arbitrary single-qubit unitary without using quantum-state teleportation. Their construction decomposes an arbitrary single-qubit unitary into elementary rotations and realizes these operations using an entangled quantum resource and classical communication. Their approach demonstrates that an arbitrary single-qubit operation can be implemented remotely while the target state remains away from the operation holder.

In the present work, we consider a different information-access scenario. In particular, we consider a blind remote-implementation setting in which Alice does not have access to the classical parameters of the operations that she physically implements. Instead, Charlie determines the required decomposition of the target operation and provides Alice with calibrated black-box devices for implementing the corresponding elementary operations. Alice can therefore perform the required operations without knowing their parameters or the classical description of the target unitary. The use of pre-shared entanglement allows the action of these operations to be implemented remotely on Bob's target qubit(s).

The key feature of the proposed framework is that the operation is implemented on Bob's original target qubit(s), rather than on a teleported copy of the quantum state at Alice's side. Bob's unknown state therefore remains at Bob's location throughout the complete protocol. In particular, for the two-qubit case, if Bob initially holds $\ket{\xi}_{T_1T_2}$, the protocol directly transforms these same target qubits into $U\ket{\xi}_{T_1T_2}$. No physical transfer of the target state to Alice is required. Thus, the proposed framework separates the physical implementation of the operation from both the classical knowledge of the operation and the location of the target quantum state.

In this work, we first construct a remote-implementation protocol for the single-qubit operation $R_z(\gamma)R_x(\beta)R_z(\alpha)$ using pre-shared Bell states and black-box devices. We then develop a protocol for the two-qubit operation $R_{zz}(\theta)$ using a pre-shared Bell pair and a locally prepared ancillary qubit. The same construction is extended to $R_{xx}(\theta)$ and $R_{yy}(\theta)$ through fixed local unitary conjugations. Finally, these building blocks are combined using the Cartan decomposition $U=(U_1\otimes U_2)R_{zz}(2c_3)R_{yy}(2c_2)R_{xx}(2c_1)(V_1\otimes V_2),$ to obtain a protocol for the remote implementation of an arbitrary two-qubit unitary operation.

\section{Preliminaries}

\subsection{Single- and two-qubit rotation operators}
We first recall the Pauli operators and the corresponding single-qubit rotation operators used throughout this work. The Pauli operators are defined as
\begin{equation}
X=\sigma_x=
\begin{pmatrix}
0 & 1\\
1 & 0
\end{pmatrix},
\qquad
Y=\sigma_y=
\begin{pmatrix}
0 & -i\\
i & 0
\end{pmatrix},
\qquad
Z=\sigma_z=
\begin{pmatrix}
1 & 0\\
0 & -1
\end{pmatrix}.
\end{equation}

The single-qubit rotation operators about the $x$, $y$, and $z$ axes are defined by
\begin{equation}
R_x(\theta)=e^{-i\theta X/2}=
\begin{pmatrix}
\cos\frac{\theta}{2} & -i\sin\frac{\theta}{2}\\
-i\sin\frac{\theta}{2} & \cos\frac{\theta}{2}
\end{pmatrix},
\end{equation}

\begin{equation}
R_y(\theta)=e^{-i\theta Y/2}=
\begin{pmatrix}
\cos\frac{\theta}{2} & -\sin\frac{\theta}{2}\\
\sin\frac{\theta}{2} & \cos\frac{\theta}{2}
\end{pmatrix},
\end{equation}
and
\begin{equation}
R_z(\theta)=e^{-i\theta Z/2}=
\begin{pmatrix}
e^{-i\theta/2} & 0\\
0 & e^{i\theta/2}
\end{pmatrix}.
\end{equation}

For two qubits, we define the corresponding two-qubit rotation operators as
\begin{equation}
R_{xx}(\theta)=e^{-i\theta(X\otimes X)/2},
\qquad
R_{yy}(\theta)=e^{-i\theta(Y\otimes Y)/2},
\qquad
R_{zz}(\theta)=e^{-i\theta(Z\otimes Z)/2}.
\end{equation}

In particular, the two-qubit operator $R_{zz}(\theta)$ has the matrix representation
\begin{equation}
R_{zz}(\theta)=
\begin{pmatrix}
e^{-i\theta/2} & 0 & 0 & 0\\
0 & e^{i\theta/2} & 0 & 0\\
0 & 0 & e^{i\theta/2} & 0\\
0 & 0 & 0 & e^{-i\theta/2}
\end{pmatrix}.
\end{equation}

\subsection{Decomposition of single-qubit and two-qubit unitary operations}
\label{sec:decomp}

An arbitrary single-qubit unitary $U\in U(2)$ can be written, up to an overall phase, as a product of three rotations about two non-parallel axes. Choosing the $z$ and $x$ axes, we have
\begin{equation}
U=R_z(\gamma)R_x(\beta)R_z(\alpha),
\label{eq_sd}
\end{equation}
where $\alpha$, $\beta$, and $\gamma$ are real parameters. This is the standard Euler decomposition of a single-qubit unitary. There are several other decompositions of a single-qubit unitary available. Recently, Shi et al. \cite{Shi2025} showed that remote implementation of a single-qubit unitary based on the above decomposition is possible.

An arbitrary two-qubit unitary operation $U\in U(4)$ can be expressed as a global phase multiplied by a special unitary operator $\widetilde{U}\in SU(4)$, i.e., $U=e^{i\phi}\widetilde{U}$, where $\phi\in\mathbb{R}$ and the global phase $e^{i\phi}$ has no physical effect on the quantum operation. Therefore, without loss of generality, the Cartan (KAK) decomposition of $SU(4)$ can be used to express an arbitrary two-qubit unitary in terms of local single-qubit unitaries and a nonlocal core as
\begin{equation}
\widetilde{U}=(U_1\otimes U_2)N(\vec{c})(V_1\otimes V_2),
\qquad
N(\vec{c})=\exp\!\left[-i\left(c_1X\otimes X+c_2Y\otimes Y+c_3Z\otimes Z\right)\right],
\label{eq_kak}
\end{equation}
where $U_1,U_2,V_1,V_2\in SU(2)$ and $c_1,c_2,c_3\in[0,\pi/4]$ \cite{Khaneja2001}. Each of the four local unitaries can in turn be expressed using the single-qubit Euler decomposition in Eq.~\eqref{eq_sd}.

Since the operators $X\otimes X$, $Y\otimes Y$, and $Z\otimes Z$ mutually commute, the nonlocal part can be factorized as
\begin{equation}
N(\vec{c})=R_{zz}(2c_3)R_{yy}(2c_2)R_{xx}(2c_1),
\qquad
R_{PP}(\theta):=\exp\!\left(-\frac{i\theta}{2}P\otimes P\right),
\label{eq:factor}
\end{equation}
where $P\in\{X,Y,Z\}$.

The two-qubit rotations $R_{xx}(\theta)$ and $R_{yy}(\theta)$ can be obtained from $R_{zz}(\theta)$ through local unitary conjugations. In particular, using the Hadamard gate $H$ and $K=R_x(\pi/2)$, we have
\begin{equation}
R_{xx}(\theta)=(H\otimes H)R_{zz}(\theta)(H\otimes H),
\label{eq:rxx_rzz}
\end{equation}
and
\begin{equation}
R_{yy}(\theta)=(K^\dagger\otimes K^\dagger)R_{zz}(\theta)(K\otimes K),
\qquad
K=R_x(\pi/2).
\label{eq:ryy_rzz}
\end{equation}

Consequently, an arbitrary two-qubit unitary can be expressed, up to an overall global phase, as
\begin{equation}
\begin{aligned}
U=&(U_1\otimes U_2)R_{zz}(2c_3)(K^\dagger\otimes K^\dagger)R_{zz}(2c_2)(K\otimes K)\\
&~~~~~~(H\otimes H)R_{zz}(2c_1)(H\otimes H)(V_1\otimes V_2).
\end{aligned}
\label{eq:u_rzz}
\end{equation}

Thus, the implementation of an arbitrary two-qubit unitary operation can be reduced to the implementation of arbitrary single-qubit unitaries of the form given in Eq.~\eqref{eq_sd} and a single type of two-qubit rotation, namely $R_{zz}(\theta)$.

\section[]{Teleportation of $R_z(\gamma)R_x(\beta)R_z(\alpha)$}
\label{sec:rzrxrz}

In this section, we first consider the remote implementation of an arbitrary single-qubit unitary $U=R_z(\gamma)R_x(\beta)R_z(\alpha)$ using pre-shared entanglement, local quantum operations, and classical communication. Bob holds a target qubit $B$ in an unknown state $\ket{\phi}_B=m\ket{0}+n\ket{1}$, where $|m|^2+|n|^2=1$, and the objective is to transform this same target qubit into $R_z(\gamma)R_x(\beta)R_z(\alpha)\ket{\phi}_B$ without transferring the unknown state to Alice. Charlie has already supplied Alice with three calibrated black-box devices, $\mathcal{C}_{\alpha}$, $\mathcal{C}_{\beta}$, and $\mathcal{C}_{\gamma}$, which implement $R_z(\alpha)$, $R_x(\beta)$, and $R_z(\gamma)$, respectively. Alice does not know the values of $\alpha$, $\beta$, and $\gamma$, nor does she have access to the classical description of the operations implemented by these devices.

Before the protocol begins, Alice and Bob share two Bell pairs, $\ket{\Phi^+}_{A_1B_1}$ and $\ket{\Phi^+}_{A_2B_2}$, where $\ket{\Phi^+}=\frac{1}{\sqrt{2}}(\ket{00}+\ket{11})$. Alice holds $A_1$ and $A_2$, while Bob holds $B_1$ and $B_2$. Thus, the initial five-qubit state is
\begin{equation}
    \begin{split}
        &\ket{\Phi^+}_{A_1B_1}\otimes \ket{\Phi^+}_{A_2B_2} \otimes\ket{\phi}_B\\
        =&\frac{1}{\sqrt{2}}(\ket{00}+\ket{11})_{A_1B_1}\otimes\frac{1}{\sqrt{2}}(\ket{00}+\ket{11})_{A_2B_2}\otimes (m\ket{0}+n\ket{1})_B.
    \end{split}
\end{equation}

The protocol proceeds as follows.

\textbf{Step 1:} Bob applies a controlled-NOT gate with $B$ as the control qubit and $B_1$ as the target qubit. He then applies another controlled-NOT gate with $B_2$ as the control qubit and $B$ as the target qubit. Consequently, the resulting state of the qubits $A_1$, $A_2$, $B_1$, $B_2$, and $B$ becomes
\begin{equation}
    \begin{split}
        \big(&m\ket{00000}+m\ket{01011}+m\ket{10100}+m\ket{11111}\\
        &+n\ket{00101}+n\ket{01110}+n\ket{10001}+n\ket{11010}\big)_{A_1A_2B_1B_2B}.
    \end{split}
\end{equation}

\textbf{Step 2:} Bob measures qubit $B_1$ in the computational basis ${\ket{0},\ket{1}}$ and obtains the classical outcome $b_1\in{0,1}$, where $b_1=0$ and $b_1=1$ correspond to the measurement outcomes $\ket{0}$ and $\ket{1}$, respectively. Conditioned on the measurement outcome, the resulting state of the qubits $A_1$, $A_2$, $B_2$, $B$ is
\begin{equation}
    \begin{split}
        \big(m\ket{b_1}\ket{000}+m\ket{b_1}\ket{111}+n\ket{1\oplus b_1}\ket{001}+n\ket{1\oplus b_1}\ket{110}\big)_{A_1A_2B_2B} 
    \end{split}
\end{equation}

\textbf{Step 3:} Bob measures qubit $B_2$ in the Hadamard basis ${\ket{+},\ket{-}}$ and obtains the classical outcome $b_2\in{0,1}$, where $b_2=0$ and $b_2=1$ correspond to the measurement outcomes $\ket{+}$ and $\ket{-}$, respectively. Then the reduced state of qubits $A_1$, $A_2$, and $B$ becomes
\begin{equation}
    \begin{split}
        \big(m\ket{b_1}\ket{00}+(-1)^{b_2}m\ket{b_1}\ket{11}+n\ket{1\oplus b_1}\ket{01}+(-1)^{b_2}n\ket{1\oplus b_1}\ket{10}\big)_{A_1A_2B}.
    \end{split}
\end{equation}

\textbf{Step 4:} Bob sends the classical bits $(b_1,b_2)$ to Alice through a classical channel.

\textbf{Step 5:} Alice performs the correction $\mathcal{U}_{Alice}$ on $(A_1,A_2)$ according to the measurement outcomes $(b_1,b_2)$, as given in Table~\ref{tab}. After applying the correction, the state of qubits $A_1$, $A_2$, $B$ becomes
\begin{equation}
    \begin{split}
        (m\ket{000}+m\ket{111}+n\ket{101}+n\ket{010})_{A_1A_2B}.
    \end{split}
\end{equation}

\textbf{Step 6:} Alice introduces an ancillary qubit $A_3$ in the state $\ket{0}$. She applies a controlled-NOT gate with $A_1$ as the control qubit and $A_3$ as the target qubit, followed by a controlled-NOT gate with $A_2$ as the control qubit and $A_3$ as the target qubit. Thereafter, the state of qubits $A_1$, $A_2$, $A_3$, $B$ is given by
\begin{equation}
    \begin{split}
        (m\ket{0000}+m\ket{1101}+n\ket{1011}+n\ket{0110})_{A_1A_2A_3B}.
    \end{split}
    \label{3.6}
\end{equation}

\textbf{Step 7:} Alice sends $A_3$ through the black-box device $\mathcal{C}_{\alpha}$, which implements $R_z(\alpha)$. She then measures $A_3$ in the Hadamard basis $\{\ket{+},\ket{-}\}$ and obtains a classical bit $a_3$, where $a_3=0$ ($1$) corresponds to the outcome $\ket{+}$ ($\ket{-}$). Subsequently, the state in Eq. \eqref{3.6} is converted to
\begin{equation}
    \begin{split}
        (me^{\frac{-i\alpha}{2}}\ket{000}+me^{\frac{-i\alpha}{2}}\ket{111}+(-1)^{a_3}ne^{\frac{i\alpha}{2}}\ket{101}+(-1)^{a_3}ne^{\frac{i\alpha}{2}}\ket{010})_{A_1A_2B}.
    \end{split}
\end{equation}

\textbf{Step 8:} According to $a_3$, Alice applies $(\sigma_z)^{a_3}$ on qubit $A_2$. She then sends $A_2$ through the black-box device $\mathcal{C}_{\beta}$, which implements $R_x(\beta)$. After the operation, Alice applies again $(\sigma_z)^{a_3}$ on $A_2$ and measures $A_2$ in the computational basis, obtaining a classical bit $a_2$. The reduced state of qubits $A_1$, $B$ is given by
\begin{equation}
    \begin{split}
        \bigg[&\{(-1)^{a_2\times a_3}me^{\frac{-i\alpha}{2}}(cos\frac{\beta}{2})^{1\oplus a_2}(-i sin\frac{\beta}{2})^{a_2}+(-1)^{a_2\times a_3}ne^{\frac{i\alpha}{2}}(-isin\frac{\beta}{2})^{1\oplus a_2}(cos\frac{\beta}{2})^{a_2}\}\ket{00}\\
        &+\{(-1)^{a_2\times a_3}(-1)^{a_3}me^{\frac{-i\alpha}{2}}(-isin\frac{\beta}{2})^{1\oplus a_2}(cos\frac{\beta}{2})^{a_2}\\
        &+(-1)^{a_2\times a_3}(-1)^{a_3}ne^{\frac{i\alpha}{2}}(cos\frac{\beta}{2})^{1\oplus a_2}(-isin\frac{\beta}{2})^{a_2}\}\ket{11}\bigg]_{A_1B}.
    \end{split}
\end{equation}

\textbf{Step 9:} Alice applies $(\sigma_x)^{a_2}$ on qubit $A_1$ based on classical bit $a_2$. She then sends $A_1$ through the black-box device $\mathcal{C}_{\gamma}$, which implements $R_z(\gamma)$. Alice applies $(\sigma_x)^{a_2}$ on $A_1$ again. She then measures $A_1$ in the Hadamard basis $\{\ket{+},\ket{-}\}$ and obtains a classical bit $a_1$, where $a_1=0$ ($1$) corresponds to the outcome $\ket{+}$ ($\ket{-}$). At this stage the reduced state of qubits $B$ is
\begin{equation}
    \begin{split}
        [(\sigma_z)^{a_3}(\sigma_x)^{a_2}(\sigma_z)^{a_1}]^{-1}\bigg[e^{\frac{-i\gamma}{2}}\{cos\frac{\beta}{2}me^{\frac{-i\alpha}{2}}-isin\frac{\beta}{2}ne^{\frac{i\alpha}{2}}\}|0\rangle+e^{\frac{i\gamma}{2}}\{-isin\frac{\beta}{2}me^{\frac{-i\alpha}{2}}+cos\frac{\beta}{2}ne^{\frac{i\alpha}{2}}\}|1\rangle\bigg].
    \end{split}
\end{equation}

\textbf{Step 10:} Alice sends the classical bits $(a_1,a_2,a_3)$ to Bob through a classical channel.

\textbf{Step 11:} Finally Bob applies a suitable unitary operation $\mathcal{U}_{Bob}=(\sigma_z)^{a_3}(\sigma_x)^{a_2}(\sigma_z)^{a_1}$ on qubit $B$ based on the classical messages $a_1$, $a_2$ and $a_3$.

After Step 11, Bob's target qubit is in the state $\ket{\phi_{\mathrm{out}}}_B=R_z(\gamma)R_x(\beta)R_z(\alpha)\ket{\phi}_B.$

\begin{table}[H]
\centering
\begin{tabular}{@{}c|l@{}}
\hline
$(b_1,b_2)$ & Alice's correction $\mathcal{U}_{Alice}$ on $(A_1,A_2)$  \\
\hline
$(0,0)$ & CNOT${A_2\rightarrow A_1}$ \\
$(0,1)$ & CNOT${A_2\rightarrow A_1}$, followed by $Z_{A_2}$ \\
$(1,0)$ & ${\sigma_x}_{A_1}$, followed by CNOT${A_2\rightarrow A_1}$ \\
$(1,1)$ & ${\sigma_x}_{A_1}$, followed by CNOT${A_2\rightarrow A_1}$, followed by ${\sigma_z}_{A_2}$ \\
\hline
\end{tabular}
\caption{Alice's correction operations according to the measurement
outcomes $(b_1,b_2)$. Here, $b_1,b_2\in\{0,1\}$.}
\label{tab}
\end{table}

Thus, the operation $R_z(\gamma)R_x(\beta)R_z(\alpha)$ is successfully implemented on Bob's unknown target qubit without transferring the state to Alice. At no stage does Alice require the classical values of $\alpha$, $\beta$, or $\gamma$; she interacts with the three operations only through the corresponding black-box devices supplied by Charlie. Since any single-qubit unitary can be decomposed, up to a global phase, as $U=R_z(\gamma)R_x(\beta)R_z(\alpha)$, the proposed protocol therefore implements an arbitrary single-qubit unitary for arbitrary values of $\alpha$, $\beta$, and $\gamma$, without requiring these parameters to be known to Alice.

\begin{figure}
    \centering
    \includegraphics[width=1\linewidth]{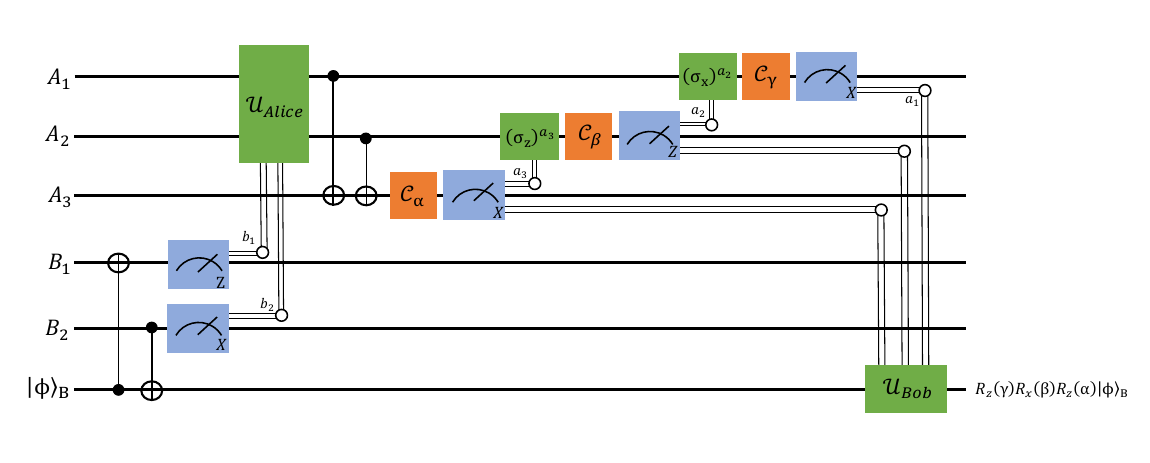}
    \caption{Quantum circuit for the remote implementation of the single-qubit unitary $U=R_z(\gamma)R_x(\beta)R_z(\alpha)$. The unknown state $\ket{\phi}_B$ remains at Bob's side throughout the protocol, while Alice implements the required operations through the black-box devices $\mathcal{C}_{\alpha}$, $\mathcal{C}_{\beta}$, and $\mathcal{C}_{\gamma}$.}
    \label{cir_sing}
\end{figure}

The quantum circuit corresponding to the above protocol is shown in Fig.~\ref{cir_sing}. The circuit illustrates the complete sequence of local operations, measurements, classical communication, and measurement-dependent corrections required to remotely implement the single-qubit unitary $U=R_z(\gamma)R_x(\beta)R_z(\alpha)$ on Bob's qubit.

\begin{figure}
    \centering
    \includegraphics[width=0.3\linewidth]{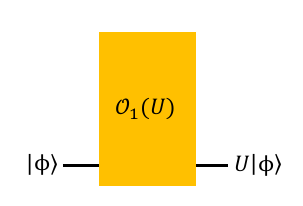}
    \caption{Effective operation block $\mathcal{O}_1(U)$ for the remote implementation of the single-qubit unitary $U=R_z(\gamma)R_x(\beta)R_z(\alpha)$ on Bob's unknown target qubit.}
    \label{single_block}
\end{figure}

For convenience, the complete remote-implementation protocol for an arbitrary single-qubit unitary $U$ is represented by the effective operation block $\mathcal{O}_1(U)$, as shown in Fig.~\ref{single_block}. Here, $\mathcal{O}_1(U)$ represents the complete protocol described above, including the pre-shared entanglement, local quantum operations, measurements, classical communication, and measurement-dependent corrections.

\section[]{Teleportation of $R_{zz}(\theta)$}
\label{sec:rzz}

In this section, we consider the remote implementation of the two-qubit unitary operation $R_{zz}(\theta)$. Suppose Bob holds two qubits $(T_1,T_2)$ in an unknown state $\ket{\psi}_{T_1T_2}=p\ket{00}+q\ket{01}+r\ket{10}+s\ket{11}$, and the goal is to transform this state into $R_{zz}(\theta)\ket{\psi}_{T_1T_2}$. As in Section~\ref{sec:rzrxrz}, Charlie supplies Alice with a black-box device $\mathcal{C}_{zz}$ that implements $R_{zz}(\theta)$ on two input qubits. Alice can use this device without knowing the value of $\theta$ or the classical description of the operation.

Before the protocol begins, Alice and Bob share one Bell pair $\ket{\Phi^+}_{AB}$, where Alice holds $A$ and Bob holds $B$. The joint state of the four qubits is
\begin{equation}
    \begin{split}
        \frac{1}{\sqrt{2}}(\ket{00}+\ket{11})_{AB}\otimes(p\ket{00}+q\ket{01}+r\ket{10}+s\ket{11})_{T_1T_2}.
    \end{split}
    \label{zz0}
\end{equation}

The protocol proceeds as follows:

\textbf{Step 1:} Bob applies a controlled-NOT gate with $T_1$ as the control qubit and $T_2$ as the target qubit. Then the state in Eq. \eqref{zz0} becomes
\begin{equation}
    \begin{split}
        \frac{1}{\sqrt{2}}(\ket{00}+\ket{11})_{AB}\otimes(p\ket{00}+q\ket{01}+r\ket{11}+s\ket{10})_{T_1T_2}.
    \end{split}
    \label{zz1}
\end{equation}

\textbf{Step 2:} Bob applies a controlled-NOT gate with $T_2$ as the control qubit and $B$ as the target qubit. After that, the state in Eq. \eqref{zz1} becomes
\begin{equation}
    \begin{split}
        \frac{1}{\sqrt{2}}\big(&p\ket{0000}+p\ket{1100}+q\ket{0001}+q\ket{1101}\\
        &+r\ket{0011}+r\ket{1111}+s\ket{0010}+s\ket{1110}\big)_{ABT_1T_2}.
    \end{split}
\end{equation}

\textbf{Step 3:} Bob measures $B$ in the computational basis $\{\ket{0},\ket{1}\}$ and obtains a classical bit $g$, where $g=0~(1)$ corresponds to the measurement outcome $\ket{0}~(\ket{1})$. The reduced state of qubits $A,T_1,T_2$ can be written as
\begin{equation}
    \begin{split}
        \frac{1}{\sqrt{2}}\big(p\ket{g}\ket{00}+q\ket{1\oplus g}\ket{01}+r\ket{1\oplus g}\ket{11}+s\ket{g}\ket{10}\big)_{AT_1T_2}.
    \end{split}
\end{equation}

\textbf{Step 4:} Bob sends $g$ to Alice through a classical channel.

\textbf{Step 5:} Alice applies $(\sigma_x)^{g}$ on $A$. After this correction, the state of $(A,T_1,T_2)$ is
\begin{equation}
    \begin{split}
        \frac{1}{\sqrt{2}}\big(p\ket{000}+q\ket{101}+r\ket{111}+s\ket{010}\big)_{AT_1T_2}.
    \end{split}
    \label{zz5}
\end{equation}

\textbf{Step 6:} Alice introduces a fresh qubit $A'$ in the state $\ket{0}$ and sends the qubits $(A,A')$ through the black-box device $\mathcal{C}_{zz}$. Since $\mathcal{C}_{zz}$ implements $R_{zz}(\theta)$. The state in Eq. \eqref{zz5} is transformed into
\begin{equation}
    \begin{split}
        \frac{1}{\sqrt{2}}\big(&pe^{-i\theta/2}\ket{0000}+qe^{i\theta/2}\ket{0101}\\
        &+re^{i\theta/2}\ket{0111}+se^{-i\theta/2}\ket{0010}\big)_{A'AT_1T_2}.
    \end{split}
\end{equation}

The qubit $A'$ remains in the state $\ket{0}$ throughout the operation.

\textbf{Step 7:} Alice measures $A$ in the Hadamard basis $\{\ket{+},\ket{-}\}$ and obtains a bit $k$, where $k=0~(1)$ corresponds to the measurement outcome $\ket{+}~(\ket{-})$. The qubit $A'$ is then discarded. The two qubits $(T_1,T_2)$ are left in the state
\begin{equation}
    \begin{split}
        \big(pe^{-i\theta/2}\ket{00}+(-1)^kqe^{i\theta/2}\ket{01}+(-1)^kre^{i\theta/2}\ket{11}+se^{-i\theta/2}\ket{10}\big)_{T_1T_2}.
    \end{split}
    \label{zz7}
\end{equation}

\textbf{Step 8:} Alice sends $k$ to Bob through a classical channel.

\textbf{Step 9:} Bob applies $(\sigma_z)^{k}$ on qubit $T_2$. The state in Eq. \eqref{zz7} then
becomes
\begin{equation}
    \begin{split}
        \big(pe^{-i\theta/2}\ket{00}+qe^{i\theta/2}\ket{01}+re^{i\theta/2}\ket{11}+se^{-i\theta/2}\ket{10}\big)_{T_1T_2}.
    \end{split}
\end{equation}

\textbf{Step 10:} Bob applies a controlled-NOT gate with $T_1$ as the control qubit and $T_2$ as the target qubit. Then the state of qubits $T_1, T_2$ becomes
\begin{equation}
    \begin{split}
        &\big(pe^{-i\theta/2}\ket{00}+qe^{i\theta/2}\ket{01}+re^{i\theta/2}\ket{10}+se^{-i\theta/2}\ket{11}\big)_{T_1T_2}\\
        =&R_{zz}(\theta)\ket{\psi}_{T_1T_2}.
    \end{split}
\end{equation}

\begin{figure}
    \centering
    \includegraphics[width=1\linewidth]{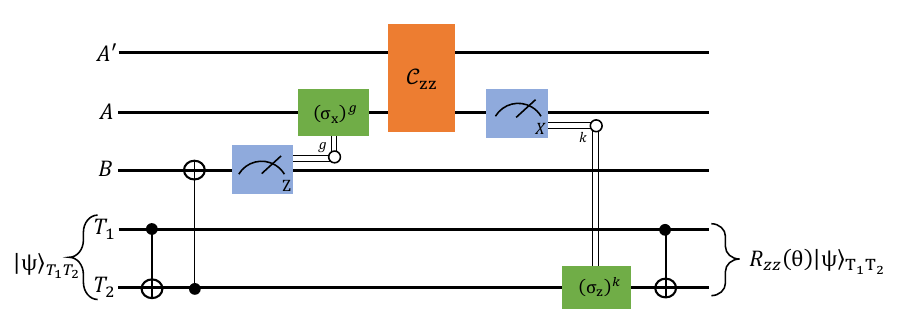}
    \caption{Quantum circuit for the remote implementation of $R_{zz}(\theta)$. Bob's unknown two-qubit state $\ket{\psi}_{T_1T_2}$ remains at his side throughout the protocol.}
    \label{fig_zz1}
\end{figure}

The quantum circuit corresponding to the remote implementation of $R_{zz}(\theta)$ is shown in Fig.~\ref{fig_zz1}. The circuit includes the pre-shared Bell state, the local operations performed by Alice and Bob, the black-box device $\mathcal{C}_{zz}$, and the measurement-dependent corrections.

\begin{figure}
    \centering
    \includegraphics[width=0.4\linewidth]{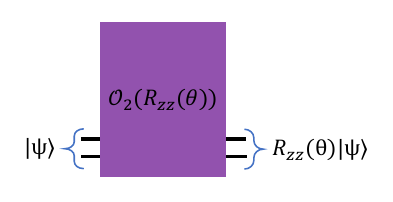}
    \caption{Effective operation block $\mathcal{O}_2(R_{zz}(\theta))$ representing the complete remote implementation of the two-qubit unitary $R_{zz}(\theta)$ on Bob's unknown target qubits.}
    \label{fig_zz2}
\end{figure}

Similarly, for convenience, the complete remote-implementation protocol for $R_{zz}(\theta)$ is represented by the effective operation block $\mathcal{O}_2(R_{zz}(\theta))$, as shown in Fig.~\ref{fig_zz2}. Here, $\mathcal{O}_2(R_{zz}(\theta))$ represents the complete protocol described above, including the pre-shared entanglement, local quantum operations, measurements, classical communication, and measurement-dependent corrections. This effective block will be used as a building block for the remote implementation of the nonlocal operations appearing in the Cartan decomposition of an arbitrary two-qubit unitary.

\subsection*{Extension to $R_{xx}(\theta)$ and $R_{yy}(\theta)$}

The $R_{zz}(\theta)$ protocol can also be used to implement $R_{xx}(\theta)$ and $R_{yy}(\theta)$ through fixed local conjugations. For $R_{xx}(\theta)$, Bob applies $H\otimes H$ before Step 1 and after Step 10. Similarly, for $R_{yy}(\theta)$, Bob applies $K\otimes K$ before Step 1 and $K^\dagger\otimes K^\dagger$ after Step 10, where $K=R_x(\pi/2)$ and $K^\dagger ZK=Y$. Thus, $$R_{xx}(\theta)=(H\otimes H)R_{zz}(\theta)(H\otimes H),$$
and $$R_{yy}(\theta)=(K^\dagger\otimes K^\dagger)R_{zz}(\theta)(K\otimes K).$$

Therefore, the same protocol can be used to implement all three nonlocal rotations remotely, with only fixed local operations added before and after the $R_{zz}(\theta)$ protocol.

\section{Teleportation of an arbitrary two-qubit unitary operation}
\label{sec:combine}

\begin{figure}
    \centering
    \includegraphics[width=1\linewidth]{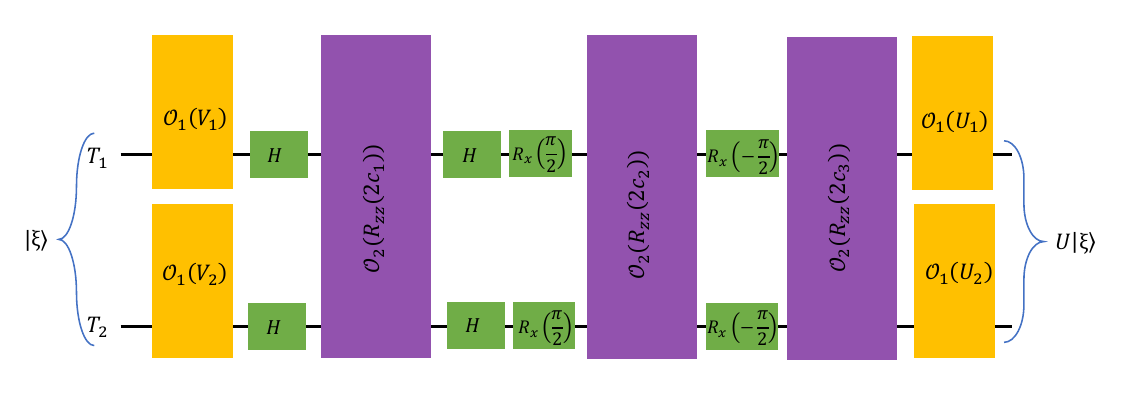}
    \caption{Compact quantum-circuit representation of the remote implementation of an arbitrary two-qubit unitary operation $U$. The blocks $\mathcal{O}_1$ and $\mathcal{O}_2$ represent the complete remote-implementation protocols for arbitrary single-qubit unitaries and $R_{zz}$-type two-qubit operations, respectively. The required $R_{xx}$ and $R_{yy}$ operations are obtained through the corresponding local conjugations.}
    \label{fig_two_qubit}
\end{figure}

We now combine the remote-implementation protocols developed in the previous sections to realize an arbitrary two-qubit unitary operation. Let Bob's target qubits $(T_1,T_2)$ be in an arbitrary unknown state $\ket{\xi}_{T_1T_2}$. The objective is to remotely implement an arbitrary two-qubit unitary operation $U\in U(4)$ on these qubits such that the initial state $\ket{\xi}_{T_1T_2}$ is transformed into $U\ket{\xi}_{T_1T_2}$. A compact representation of the complete remote-implementation protocol is shown in Fig. \ref{fig_two_qubit}.

The protocols developed in Sections~\ref{sec:rzrxrz} and \ref{sec:rzz} can now be combined with the decomposition introduced in Section~\ref{sec:decomp}. Using Eq.~\eqref{eq_kak}, an arbitrary two-qubit unitary can be expressed as
\begin{equation}
U=(U_1\otimes U_2)R_{zz}(2c_3)R_{yy}(2c_2)R_{xx}(2c_1)(V_1\otimes V_2).
\label{eq:u_kak}
\end{equation}
Thus, the complete protocol is constructed from the effective single-qubit operation block $\mathcal{O}_1(U)$ and the effective two-qubit operation block $\mathcal{O}_2(R_{zz}(\theta))$ introduced in the previous sections. The complete protocol consists of seven stages.

\textbf{Step 1:} The single-qubit remote-implementation protocol described in Section~\ref{sec:rzrxrz} is applied to Bob's qubit $T_1$ by taking $B=T_1$ and choosing the target unitary as $U=V_1$. Thus, the effective operation block $\mathcal{O}_1(V_1)$ remotely implements $V_1$ on $T_1$.

\textbf{Step 2:} Similarly, the same protocol is applied to Bob's qubit $T_2$ by taking $B=T_2$ and choosing the target unitary as $U=V_2$. Hence, the effective operation block $\mathcal{O}_1(V_2)$ remotely implements $V_2$ on $T_2$. Since the two protocols act on different target qubits, Steps 1 and 2 can be performed in parallel. After their completion, the combined operation on $(T_1,T_2)$ is $V_1\otimes V_2.$

\textbf{Step 3:} To remotely implement $R_{xx}(2c_1)$, the two-qubit remote-implementation protocol of Section~\ref{sec:rzz} is applied to the target qubits $(T_1,T_2)$ with $\theta=2c_1$. First, the local operation $H\otimes H$ is applied to $(T_1,T_2)$. The effective operation block $\mathcal{O}_2(R_{zz}(2c_1))$ is then applied, followed by another local operation $H\otimes H$. Therefore, the overall operation implemented in this stage is $$(H\otimes H)R_{zz}(2c_1)(H\otimes H)=R_{xx}(2c_1).$$

\textbf{Step 4:} To remotely implement $R_{yy}(2c_2)$, the two-qubit remote-implementation protocol of Section~\ref{sec:rzz} is applied to $(T_1,T_2)$ with $\theta=2c_2$. First, the local operation $K\otimes K$ is applied to the target qubits, where $K=R_x(\pi/2)$. The effective operation block $\mathcal{O}_2(R_{zz}(2c_2))$ is then applied, followed by the local operation $K^\dagger\otimes K^\dagger$. Thus, the overall operation implemented in this stage is $$(K^\dagger\otimes K^\dagger)R_{zz}(2c_2)(K\otimes K)=R_{yy}(2c_2).$$

\textbf{Step 5:} The two-qubit remote-implementation protocol of Section~\ref{sec:rzz} is applied directly to $(T_1,T_2)$ with $\theta=2c_3$. No local conjugation is required in this case. Therefore, the effective operation block $\mathcal{O}_2(R_{zz}(2c_3))$ directly implements$ R_{zz}(2c_3)$.

Steps 3--5 are performed sequentially because they act on the same target qubits $(T_1,T_2)$. Their chronological order is $R_{xx}(2c_1)$, followed by $R_{yy}(2c_2)$ and then $R_{zz}(2c_3)$, which gives the nonlocal part
\begin{equation}
R_{zz}(2c_3)R_{yy}(2c_2)R_{xx}(2c_1)
\end{equation}
of the Cartan decomposition.

\textbf{Step 6:} The single-qubit remote-implementation protocol of Section~\ref{sec:rzrxrz} is applied to Bob's qubit $T_1$ by taking $B=T_1$ and choosing the target unitary as $U=U_1$. Thus, the effective operation block $\mathcal{O}_1(U_1)$ remotely implements $U_1$ on $T_1$.

\textbf{Step 7:} Similarly, the single-qubit protocol is applied to Bob's qubit $T_2$ by taking $B=T_2$ and choosing the target unitary as $U=U_2$. Thus, the effective operation block $\mathcal{O}_1(U_2)$ remotely implements $U_2$ on $T_2$. Since Steps 6 and 7 act on different target qubits, they can be performed in parallel. After their completion, the combined operation is $U_1\otimes U_2$.

Therefore, the complete sequence of operations shown in Fig.~\ref{fig_two_qubit} is
\begin{equation}
V_1\otimes V_2\longrightarrow R_{xx}(2c_1)\longrightarrow R_{yy}(2c_2)\longrightarrow R_{zz}(2c_3)\longrightarrow U_1\otimes U_2.
\end{equation}
This sequence realizes the Cartan decomposition of the arbitrary two-qubit unitary $U$. Consequently, Bob's initial state $\ket{\xi}_{T_1T_2}$ is transformed into $\ket{\xi_{\mathrm{out}}}_{T_1T_2}=U\ket{\xi}_{T_1T_2}$.

This completes the proposed protocol for the remote implementation of an arbitrary two-qubit unitary operation.

\section{Discussion and Conclusion}
In this work, we have developed a framework for the blind remote implementation of an arbitrary two-qubit unitary operation. We first constructed a remote-implementation protocol for the single-qubit operation $R_z(\gamma)R_x(\beta)R_z(\alpha)$ and then developed a protocol for the two-qubit operation $R_{zz}(\theta)$. By using fixed local conjugations, the same $R_{zz}$ protocol is extended to $R_{xx}(\theta)$ and $R_{yy}(\theta)$. These building blocks are finally combined through the Cartan decomposition to realize an arbitrary two-qubit unitary $U$ on Bob's unknown state $\ket{\xi}_{T_1T_2}$.

The complete remote implementation of an arbitrary two-qubit unitary is constructed from four instances of the single-qubit protocol $\mathcal{O}_1$ and three instances of the two-qubit protocol $\mathcal{O}_2$. Each $\mathcal{O}_1$ requires two pre-shared Bell states and five classical bits, whereas each $\mathcal{O}_2$ requires one pre-shared Bell state and two classical bits. Therefore, the complete protocol requires eleven Bell states and twenty-six classical bits. In addition, Alice requires the corresponding calibrated black-box devices for the elementary operations appearing in the decomposition of $U$.

\vspace{2em}

\noindent\textbf{Acknowledgments:} This research work of the first author is supported by the University Grants Commission (UGC), Government of India (NTA Ref. No.: 211610140697, dated: 19th April 2022). The valuable suggestions of the reviewers are gratefully acknowledged.\\

\noindent\textbf{Data availability statement:} Our manuscript has no associated data.\\

\noindent\textbf{Conflict of interest:} On behalf of all authors, the corresponding author states that there is no conflict of interest.\\

\end{document}